# Transmission function properties for multi-layered structures: Application to super-resolution

N. Mattiucci,[1] G. D'Aguanno,[1*] M. Scalora,[1] M. J. Bloemer[1] and C. Sibilia[2]

[1] *Dept. of the Army. Charles M. Bowden Facility, Bldg 7804, Research Development and Engineering Command, Redstone Arsenal, AL 35898, USA*
[2] *Dipartimento di Energetica, Università "La Sapienza" di Roma, Via Scarpa 16, 00161, Roma, Italy*
[*] *giuseppe.daguanno@us.army.mil*

**Abstract**

We discuss the properties of the transmission function in the *k*-space for a generic multi-layered structure. In particular we analytically demonstrate that a transmission greater than one in the evanescent spectrum (amplification of the evanescent modes) can be directly linked to the guided modes supported by the structure. Moreover we show that the slope of the phase of the transmission function in the propagating spectrum is inversely proportional to the ability of the structure to compensate the diffraction of the propagating modes. We apply these findings to discuss several examples where super-resolution is achieved thanks to the simultaneous availability of the amplification of the evanescent modes and the diffraction compensation of the propagating modes.

**1. Introduction**

As first noted by Pendry [1], the amplification of evanescent waves by an ideal negative index material (NIM) layer together with the negative refraction can lead to super-resolution, i.e. resolution below the Abbe limit [2]. This theoretical prediction was later confirmed by experiments carried out using a flat lens composed of a single silver layer 40-50nm thick [3]. This phenomenon works for TM polarized light and is based on two features: negative refraction of the Poynting vector which focuses the propagating waves and excitation of surface plasmons to extend the range of the evanescent components of the source.

1-D, metallo-dielectric, multi-layered structures [4-5] have been considered in order to improve super-resolution in the near-field zone relative to the single metal layer. These multi-layered lenses do not focus the propagating waves but instead channel the light in the forward direction thanks to the negative refraction in the metallic layers and the positive refraction in the dielectric layers. Therefore, there are two equally important characteristics that contribute to achieve super-resolution in multi-layered metallo-dielectric structures. The first one is that for TM polarization they are able to compensate the diffraction of the propagating modes through a canalization mechanism [6]. The second characteristic is that the lens is able to reconstruct or amplify the near field components of the object giving therefore the possibility to beat the Abbe criterion [7]. For the metallo-dielectric super-resolving lens it is worth to mention the important role played by the thickness of the metallic layers. For thick metallic layers, say 50nm and beyond, only the evanescent part of the spectrum emanated from an object can be efficiently coupled with the lens via plasmonic guided modes, whereas for thinner metallic layers (from ~20nm to ~30nm), in addition to the coupling of the evanescent part of the spectrum, resonant tunneling of the propagating part of the spectrum may occur giving the super-resolving lens an high degree of transparency when it is properly designed [5,6].

It is well known that in general for any planar structure the spatial filtering properties and ultimately the image formation are completely characterized by its k-space transmission function $t(k_x)$ through the angular spectrum representation technique[8]. Here we indicate by $z$ the axis orthogonal to the object (propagation direction) and by $x$ the direction parallel to the object (transverse direction) and $k_x$ the transverse wave-vector.

The aim of this paper is twofold: First, we analytically prove in Section 2 that for a generic stratified structure, included but not limited to metallo-dielectric structures, the amplification of the evanescent modes, i.e $|t(k_x)|>1$ for $k_x>k_0$ ($k_0$ is the vacuum wave-vector), is nothing else than the signature of guided modes along the transverse direction of the structure. Second, we demonstrate in Section 3 that the slope of the phase of the transmission function is inversely proportional to the ability of the structure to compensate the diffraction of the propagating modes (canalization mechanism). We apply these findings to discuss several examples where super-resolution is achieved thanks to the simultaneous availability of the amplification of the evanescent modes and the diffraction compensation of the propagating modes.

## 2. Guided Modes and Amplification of the Evanescent Modes

Let us first show that the transmission peaks found in the evanescent zone of the spectrum are associated with the presence of guiding modes, propagating in the direction parallel to the multilayer planes. In our notation we indicate with $z$ the propagation direction, i.e. the axis orthogonal to the layered structure, and with x the transverse direction, i.e. the axis parallel to the layers. Using a transfer matrix approach, the values of the H field for TM polarization at the input *(z=0)* and output *(z=L)* of the structure are linked by the following matrix relation [9]:

$$\begin{pmatrix} H_L \\ \left(\frac{1}{\varepsilon}\frac{dH}{dz}\right)_L \end{pmatrix} = \begin{pmatrix} m_{11} & m_{12} \\ m_{21} & m_{22} \end{pmatrix} \begin{pmatrix} H_0 \\ \left(\frac{1}{\varepsilon}\frac{dH}{dz}\right)_0 \end{pmatrix}, \qquad (1)$$

where $m_{11}$, $m_{12}$, $m_{21}$, $m_{22}$, are the elements of the transfer matrix of the structure. The transfer matrix of a N-layer structure depends on the thickness ($d_i$), the dielectric constant ($\varepsilon_i$), and index of refraction ($n_i$) of each layer as well as on the frequency $\omega$ and the longitudinal component of the wave vector $k_x$ in the form:

$$\begin{pmatrix} m_{11} & m_{12} \\ m_{21} & m_{22} \end{pmatrix} = \prod_{i=1}^{N} \begin{pmatrix} \cos\left(\sqrt{(n_i \frac{\omega}{c})^2 - k_x^2} d_i\right) & \frac{\varepsilon_i}{\sqrt{(n_i \frac{\omega}{c})^2 - k_x^2}} \sin\left(\sqrt{(n_i \frac{\omega}{c})^2 - k_x^2} d_i\right) \\ -\frac{\sqrt{(n_i \frac{\omega}{c})^2 - k_x^2}}{\varepsilon_i} \sin\left(\sqrt{(n_i \frac{\omega}{c})^2 - k_x^2} d_i\right) & \cos\left(\sqrt{(n_i \frac{\omega}{c})^2 - k_x^2} d_i\right) \end{pmatrix}. \quad (2)$$

The guiding condition requires that the following equation must be satisfied:

$$i\left(\frac{\sqrt{(n_0 \frac{\omega}{c})^2 - k_x^2}}{\varepsilon_0} m_{22} + \frac{\sqrt{(n_{N+1} \frac{\omega}{c})^2 - k_x^2}}{\varepsilon_{N+1}} m_{11}\right) + \left(\frac{\sqrt{(n_0 \frac{\omega}{c})^2 - k_x^2}}{\varepsilon_0} \frac{\sqrt{(n_{N+1} \frac{\omega}{c})^2 - k_x^2}}{\varepsilon_{N+1}} m_{12} - m_{21}\right) = 0, \quad (3)$$

where $n_0$ and $n_{N+1}$ are respectively the refractive index of the input and output medium. On the other hand the transmission of the structure $t(k_x)$ takes the following expression:

$$t = \frac{2i \frac{\sqrt{(n_0 \frac{\omega}{c})^2 - k_x^2}}{\varepsilon_0}}{i\left(\frac{\sqrt{(n_0 \frac{\omega}{c})^2 - k_x^2}}{\varepsilon_0} m_{22} + \frac{\sqrt{(n_{N+1} \frac{\omega}{c})^2 - k_x^2}}{\varepsilon_{N+1}} m_{11}\right) + \left(\frac{\sqrt{(n_0 \frac{\omega}{c})^2 - k_x^2}}{\varepsilon_0} \frac{\sqrt{(n_{N+1} \frac{\omega}{c})^2 - k_x^2}}{\varepsilon_{N+1}} m_{12} - m_{21}\right)}. \quad (4)$$

When the condition (3) is satisfied, the transmission goes to infinity: the guiding modes of the multilayer correspond to poles of the transmission function in the evanescent region. Strictly speaking this is the condition for true guided modes with an infinite lifetime or no losses. While the connection between the poles of the transmission and the guided modes of a generic waveguide is in itself not new [10], here our aim is to show that the so-called "amplification of the evanescent modes" is nothing more than the coupling of the guided or quasi-guided modes in a *lossy* waveguide. If losses are present the multilayer becomes a leaky waveguide. In this latter case it is no longer possible to find true guided modes of the structure, but quasi-guided modes with a finite lifetime depending on the amount of the losses. As an example, let us consider in Fig.1 a 4-period structure. The multilayer is based on two (magnetically non active: $\mu=1$) materials, one of dielectric constant $\varepsilon_1=12$, the other $\varepsilon_2=-12$. In figure 1a) we plot the dispersion curve of the structure, i.e. the solution of eq. (3) on the $k_x$ $\lambda$ plane, for $k_x/k_0>1$ (the evanescent part of the spectrum) where $k_0=2\pi/\lambda$ is the vacuum wave-vector. While, in figure 1b) we plot $log_{10}(T+1)$ ($T=|t(k_x)|^2$ is the transmittance of the structure) of the evanescent modes as a function of $k_x$ and $\lambda$. We plot $log_{10}(T+1)$ instead of plotting directly $T$ to avoid numerical problems with the poles of the transmission function in the case of no absorption. Of course, the use of the logarithmic function does not change the position of the poles or the maxima of the transmittance. In figure 1c) and 1d) we show how the absorption perturbs the system.

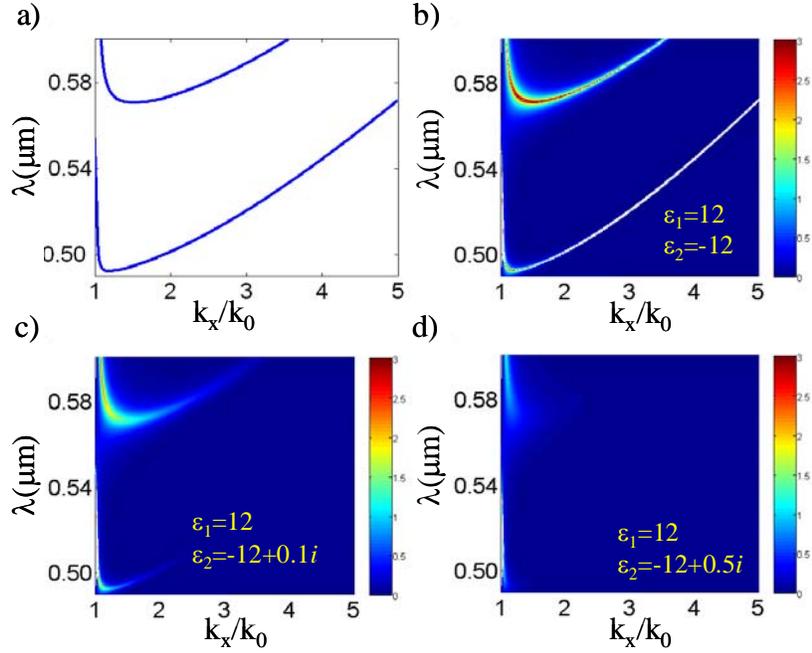

Fig. 1. a) TM dispersion curves for the guided modes of a 4 period structure $(d_1 d_2)^4$ with $d_1=d_2=40$nm. b) $\log_{10}(T+1)$ of the evanescent components vs. $\lambda$ and $k_x/k_0$ with no absorption. c) Same as b) but with small absorption. d) Same as b) but with consistent absorption.

In Fig.2 is shown that the introduction of a small absorption does not change the dispersion curve. The position of the peaks of transmission is unchanged until there is at least one order of magnitude between the real and the imaginary part of the permittivity.

On the other hand the amplitude of the peaks reduces with absorption. Instead, a small modification of the real part of the dielectric constant produces a shift of the dispersion curve.

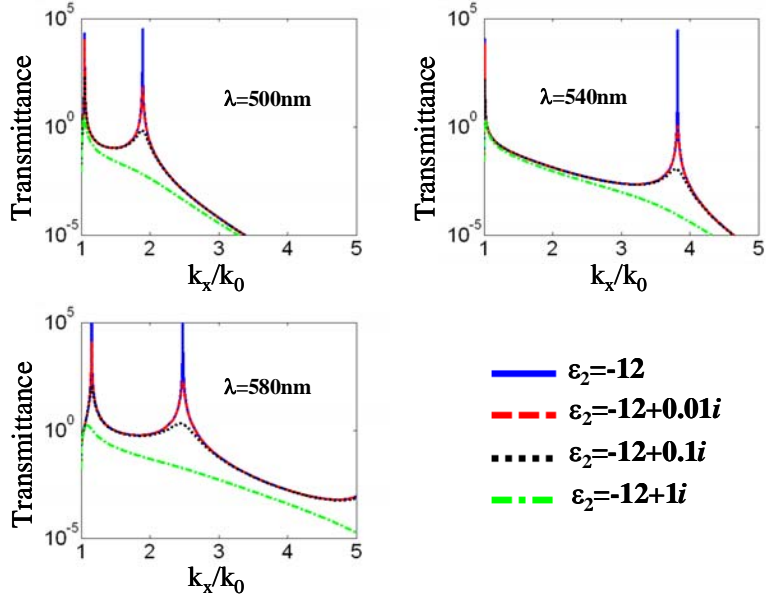

Fig. 2. Transmittance spectrum for propagating modes ($k_x/k_0<1$) and evanescent modes($k_x/k_0>1$) at different wavelengths and different values of the absorption.

These findings may be used in order to control the transmission band of the evanescent modes just by appropriately tailoring the real and imaginary part of the dielectric constants so to have a leaky mode with the desired lifetime in analogy with the theory of quasi-normal modes in leaky waveguides and, more generally, with the theory of quasi-normal modes in open systems [11].

As a last important remark before closing this Section, we would like to point out that the relationship between the amplification of the evanescent modes and the guiding properties of the multilayer structure is not a peculiar characteristic of the TM polarization. The same kind of relation holds also for TE polarization. In fact Eqs (1-4) can be rewritten as well for TE polarization with the simple substitutions: $H \to E$ and $\varepsilon \to \mu$. In Fig.3 we show the guided modes and amplification of the evanescent modes for the same structure described in Fig..1 except that this time the field is TE polarized. In other words our demonstration shows unambiguously that, as already partially remarked in Ref.[7] and Ref.[12], the amplification of evanescent modes is indeed possible in any kind of planar structure that supports guided modes, be it plasmonic or not.

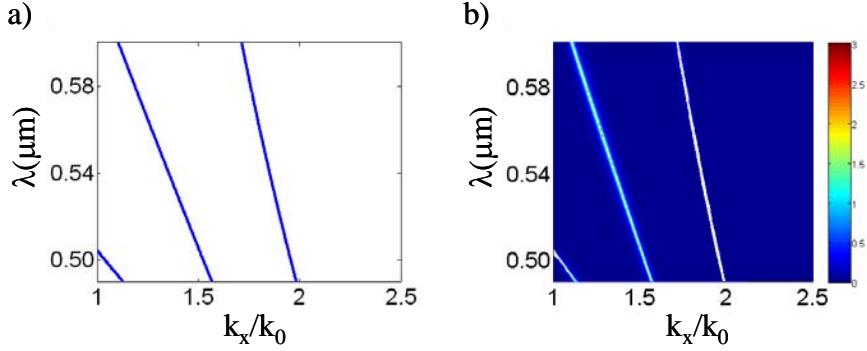

Fig. 3. a) Dispersion curves for the guided modes as in Fig.1(a) but for TE polarization. b) $\log_{10}(T+1)$ for the evanescent modes vs. $\lambda$ and $k_x/k_0$ in the case of no absorption.

Of course, amplification of the evanescent modes is just one of the two key elements that concur to achieve super-resolution, the second one is the diffraction compensation of the propagating modes (canalization mechanism). In the next Section we show that the information about the ability of the planar structure to compensate the diffraction is still contained in the transmission function of the structure, in particular in the phase of the transmission function for the propagating part of the spectrum. We will also remark that, of course, differently from the amplification of the evanescent modes, the compensation of the diffraction is achieved through the canalization mechanism only for TM polarization.

## 3. Phase of the Transmission and the Canalization Mechanism

The canalization process is intimately connected to the ability of a metallo-dielectric lens to super-resolve. This mechanism has been described in several papers [4-7, 13-14]. One pictorial way to illustrate the concept is as follows: the focusing mechanism of a metallo-dielectric multilayer lens is based on the balance between the negative refraction for TM polarization of the Poynting vector in the metal layers and the positive refraction in the dielectric layers which ultimately leads to a zig-zag path and to the "canalization" of the electromagnetic field. Nonetheless, it seems to be still unclear whether or not the canalization mechanism is itself linked to the appearance of plasmonic modes. In this Section we will give clear evidence that the canalization mechanism is associated with the phase of the transmission function of the propagation modes (the flatter the phase the better the canalization) and therefore is independent of the appearance, or lack thereof, of plasmonic modes. This means that *per se* the canalization mechanism, although a key element of the super-resolution, does not epitomize it, or we may also say it is not a synonym of super-resolution.

Let us start our analysis by considering the free-space diffraction from four slits located at z=0, as described in the caption of Fig.4. We use a scattering object made of 4 slits: the first two are 50nm wide and with a distance center to center of 150nm, the second two are 250nm wide and with a distance center to center of 600nm. We suppose a normal incident electromagnetic radiation at $\lambda=532$nm. The quantity represented in the figure is the z-component of the Poynting vector ($S_z(x,z)$) in arbitrary units. The scattering calculations have been performed using the angular spectrum representation technique [8] in conjunction with a transfer matrix technique [9]. The same calculation method has been successfully applied to describe the onset of optical vortices during a super-resolution process [15], for example.

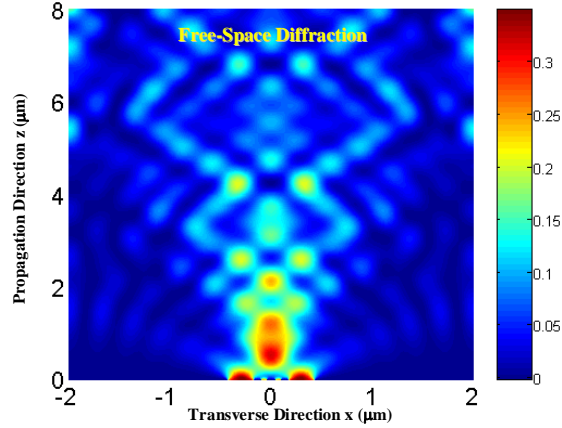

Fig. 4. Free space diffraction of the z-component of the Poynting vector in arbitrary units from 4 slits located at z=0. The 4 slits are symmetrically located with respect to the axis x=0. The two inner slits are 50nm wide and have a distance center to center of 150nm. The two outer slits are 250nm wide and have a distance center to center of 600nm.

Now we consider a metallo-dielectric lens placed in front of the scattering object. The lens and the geometry considered are detailed in Fig.5.

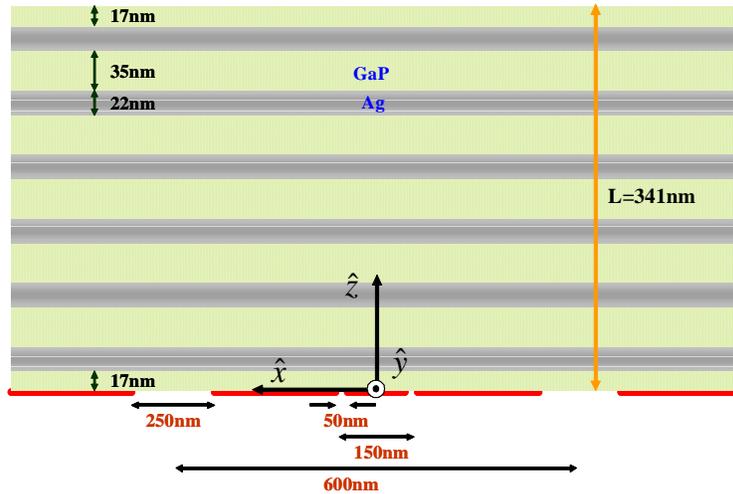

Fig. 5. Schematic representation of the lens and the slits geometry. The operative wavelength is $\lambda=532$nm. The lens is placed in vacuo.

The lens consists of 5.5 periods of Ag/GaP (22 nm/35 nm) with GaP antireflection coatings, 17 nm thick, on the entrance and exit faces, as also detailed in Ref. [5-6]. The total length of the lens is L=341nm. The permittivities for the GaP and Ag layers have been taken from experimentally measured data [16] and they are respectively: $\varepsilon_{Ag}$(532nm)=-10.17+i0.82 and $\varepsilon_{GaP}$(532nm)=12.23+i0.00367. Clearly at the output of the lens the resolution of the first two slits (the ones at a distance less than $\lambda/2$) would imply super-resolution while the resolution of the second two slits (the ones at a distance greater than $\lambda$) would imply just conventional resolution.

In Figs. 6a) and 6b) we show *$log_{10}(T+1)$* vs. *$\lambda$* and *$k_x/k_0$* respectively for TM and TE. The figures clearly show that for both polarization the lens admits quasi-guided modes although plasmonic in nature for TM polarization, while just conventional quasi-guided modes for TE polarization.

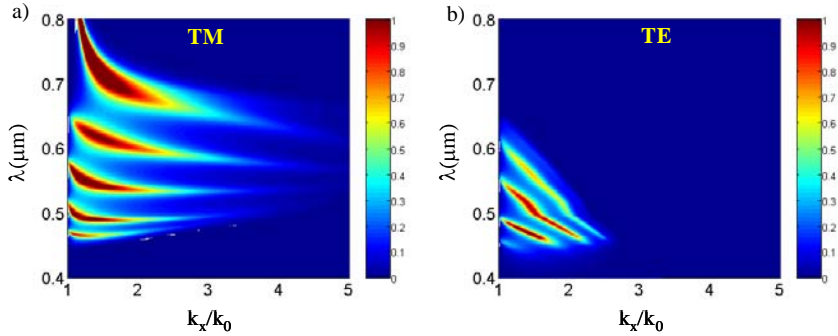

Fig. 6. $log_{10}(T+1)$ vs. $\lambda$ and $k_x/k_0$ for the Ag/GaP lens described in Fig.5. a) TM polarization. b) TE polarization.

In Fig.7a) we show the diffraction in the near field from the scattering object with no lens in front of it. In Fig.7b) and 7c) we show the diffraction from the scattering object with the lens placed in front of it, respectively for TE and TM polarization.

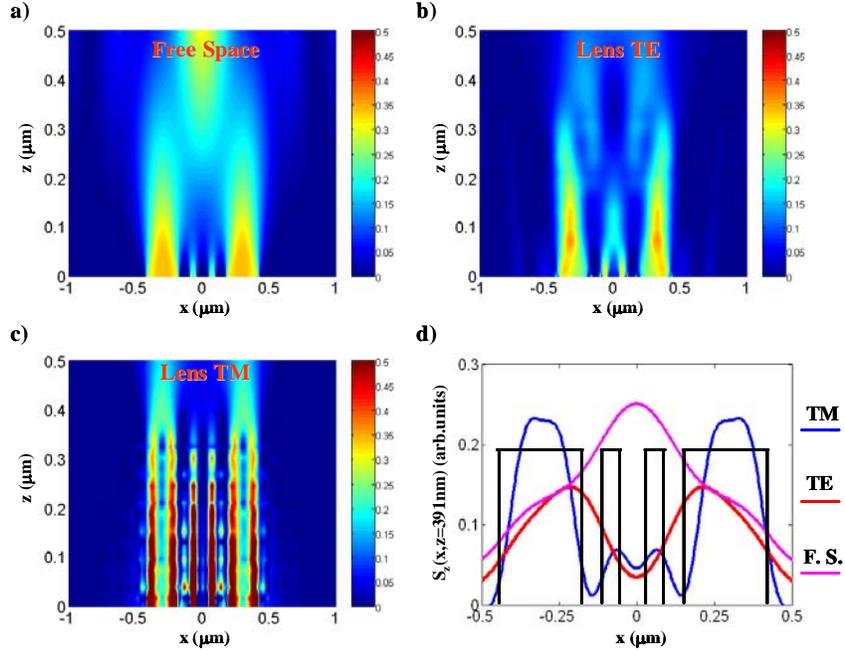

Fig. 7. a) Free space diffraction of $S_z$ from the 4 slits. b) Diffraction with the lens in front of the scattering object for TE polarization. c) Diffraction for TM polarization. d) Section of $S_z$ at z=391nm (i.e. 50nm beyond the end of the lens). Superimposed (black line) the position of the 4 slits.

Note that the details of the sub-wavelength slits are lost quite rapidly for TE polarization, while four light channels (corresponding to the canalization of the fields emitted by the 4 slits) are clearly visible for TM polarization. Note also that the canalization mechanism for the two slits at super-wavelength distance leads to strong power concentration along the lines corresponding to the edges of the 2 slits. Finally in Fig 7d) we show the section of the z-component of the Poynting vector taken at z=391nm (i.e. 50nm away from the output of the lens) for the three cases. It is evident that the two slits at subwavelength distance are resolved just for TM polarization.

Now we ask: What happens to the canalization process if the evanescent modes are removed from the spectrum of the scattering object?

In Fig.8 we show that the canalization process is still taking place quite effectively for TM polarization in the case of the two super-wavelength slits. Of course, in this case the sub-wavlength slits cannot be "seen" by the electromagnetic field, given the fact that we have artificially removed the evanescent modes from the spectrum of the scattering object. In particular in Fig.8c) we show a section of $S_z$ at z=500nm where it is evident that the two super-wavelength slits are resolved just for TM polarization. In other words the results of Fig.8 clearly suggest that the canalization process is independent of the plasmon mode excitation, in fact the canalization is taking place for TM polarization in the case of the super-wavelength slits even when the evanescent modes are removed from the spectrum, this last thing implying that no plasmonic modes can be excited.

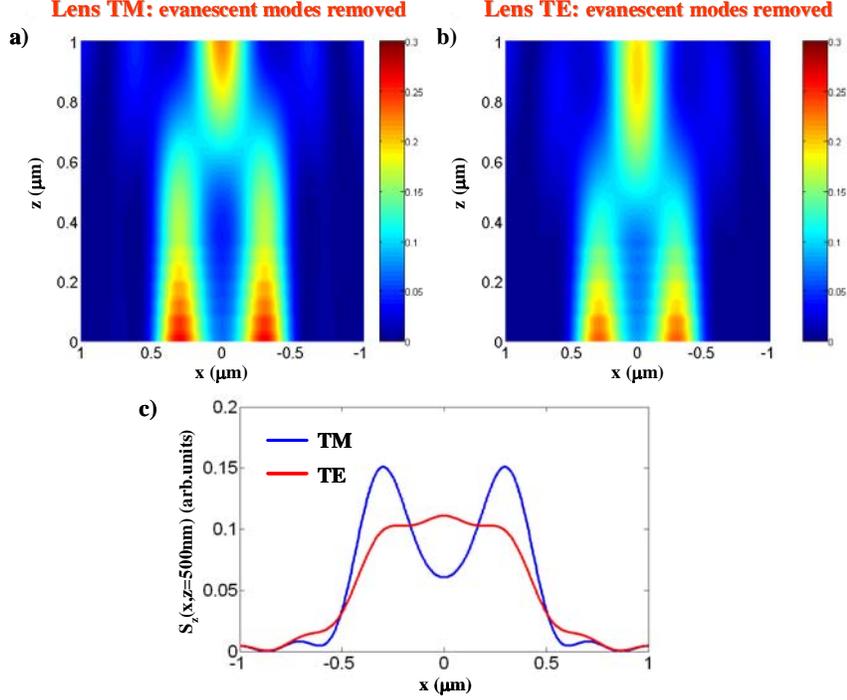

Fig. 8. a) Diffraction for TM polarization but with the evanescent modes removed from the spectrum of the scattering object. b) Same but for TE polarization. c) Section of $S_z$ at z=500nm.

Finally, in Fig.9 we show the transmittance (Fig. 9a)), i.e. $|t(k_x)|^2$, and the phase of the transmission, i.e. $\varphi(k_x)$, where $t(k_x)=|t(k_x)|exp(i\varphi(k_x))$ is the complex transmission of the structure. From Fig. 9b) we note that the best compensation of the diffraction of the propagating modes, achieved for TM polarization, corresponds to a much flatter phase in the propagation region. In other words the ability of the lens to compensate the diffraction of the propagation modes is directly proportional to the flatness of the phase of the transmission in the propagation region.

This heuristic finding is actually susceptible of a precise mathematical description. In general the phase of the transmission is an even function of $k_x$ and therefore by expanding it in a Taylor series up to the second order we obtain:

$$\varphi(k_x) = \varphi_0 + \frac{1}{2}\varphi''(k_x = 0)k_x^2 + ... \quad . \qquad (5)$$

In the simple case of the free space at a distance $z=L$ from the scattering object the phase of the transmission is:

$$\varphi_{F.S.}(k_x) = \sqrt{k_0^2 - k_x^2}\, L = k_0 L - \frac{L}{2k_0} k_x^2 + \ldots, \quad (6)$$

where clearly $\left|\varphi''_{F.S.}(k_x = 0)\right| = L/k_0 \propto L/L_D^{F.S.}$ being $L_D^{F.S.}$ the Rayleigh diffraction length of the free space [17], that is defined as $L_D^{F.S.} = k_0 w_0^2$ where $w_0$ is the beam waist or, as in our case, the slit width.

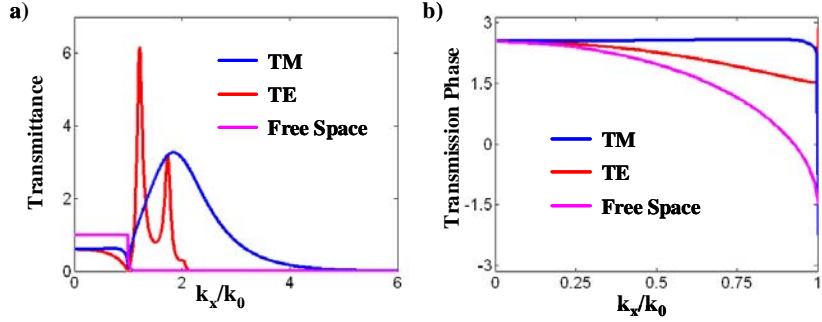

Fig. 9. a) Transmittance of the lens at λ=532nm vs. $k_x/k_0$ for propagation modes ($k_x/k_0<1$) and evanescent modes ($k_x/k_0>1$). b) Phase of the transmission for the propagation modes.

In analogy with the free space case, the second derivative of the phase of the transmission at $k_x=0$ for a generic planar structure will also be proportional to the inverse of its Rayleigh diffraction length and therefore we may write:

$$\frac{L_D}{L_D^{F.S.}} = \frac{L}{\left|\varphi''(k_x = 0)\right| k_0}, \quad (7)$$

from which it is evident that the flatter is the phase of the transmission, the better is the compensation of the diffraction of the propagating modes and therefore the better is the canalization. In our case we have that $L_D^{TE}/L_D^{F.S.} \cong 1.8$ and $L_D^{TM}/L_D^{F.S.} \cong 18$ which confirm the results shown in Fig.8, i.e. the canalization mechanism is taking place in a much more effective way only for TM polarization.

In addition to the flatness of the phase of the transmission function, the canalization mechanism can also be inferred from the equifrequency curves of the Bloch band, as also suggested in Ref.[18.]. The group velocity is perpendicular to the equifrequency curves [19], so that a flat curve over the propagation region means canalization. The standard Bloch theory applies to non absorbing materials, i.e. Hermitian systems. When dealing with real materials one can still define a generalized Bloch vector by simply extending the definition given in Ref [19] to materials with complex permittivities. Of course, in this case the Bloch vector becomes a complex quantity also outside the gap. In Fig.10 we plot the equifrequency curve for the elementary cell of our structure at λ=532nm and we compare it with the equifrequency curve of free space. The equifrequencuy curves of the elementary cell are shown for both TE and TM polarization.

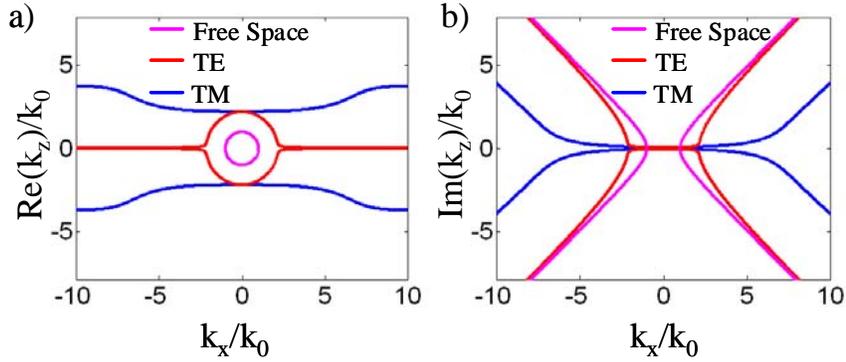

Fig. 10. Isofrequency curves at λ=532nm. a) Real part of the Bloch vector. b) Imaginary part of the Bloch vector.

Fig.10a) is the real part of the Bloch vector and Fig.10b) is the imaginary part of the Bloch vector. As one may expect, the real part of the equifrequency curve of the metallo-dilectric structure (Fig.10a)) is flat for TM polarization in the entire propagation region, an indication of canalization. Note that while the sense of the transmission function is applicable to any generic structure, the meaning of the Bloch vector when strongly absorbing materials are involved has not yet completely clarified.

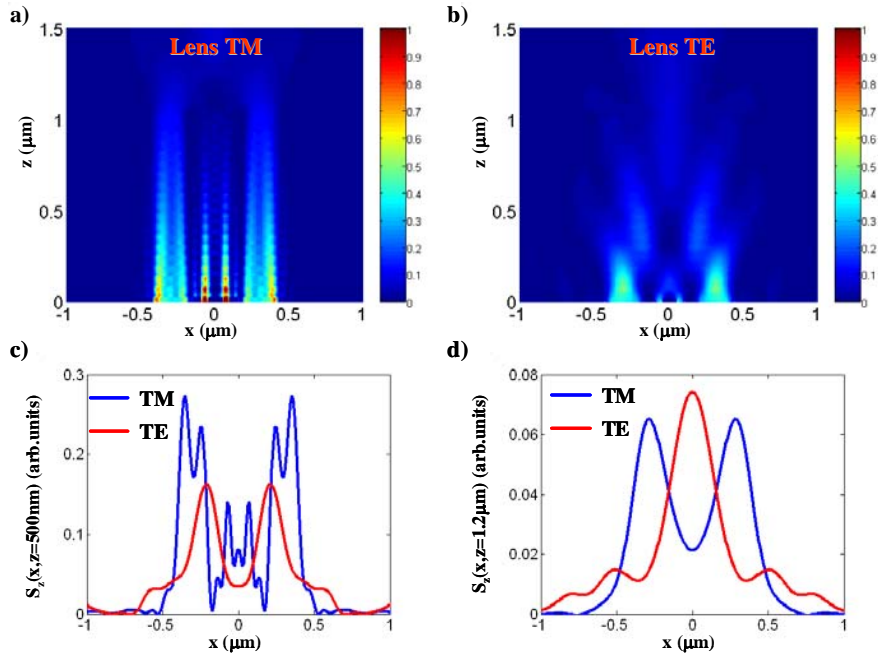

Fig. 11. Diffraction with a metallo-dielctric lens whose number of periods is triple with respect to the lens described in Fig.5. a) TM polarization. b) TE

polarization. c) Section of $S_z$ at z=500nm, i.e. in the middle of the lens. d) Section of $S_z$ at z=1200nm, i.e. ~180nm beyond the output surface of the lens.

In order to show the robustness of the canalization mechanism against the total length of the lens, in Fig.11 we show the diffraction from the 4 slits with a lens placed in front of it.

The lens is as the one described in Fig.5 except that this time the number of periods is triple, i.e. the lens total length is L=1023nm. At the output surface of the lens the two super-wavelength slits are still clearly resolved for TM polarization while no resolution is achieved for TE polarization. Note that in this case the Rayleigh diffraction length for the super-wavelength slits in free space would be $L_D^{F.S.} = k_0 w_0^2 \cong 738 nm$ ($w_0$ is the width of the slit) and the slits are resolved at the output of the lens at a distance with respect to the scattering object that is almost double the Rayleigh diffraction length of the free space. We have also calculated the diffraction from a lens with number of periods *eight times greater* with respect to the lens of Fig.5. In this case the total length of the lens is ~2.7μm. As shown in Fig. 12, we find that the canalization mechanism of the super-wavelength slits still persists which is a testament to its sturdiness.

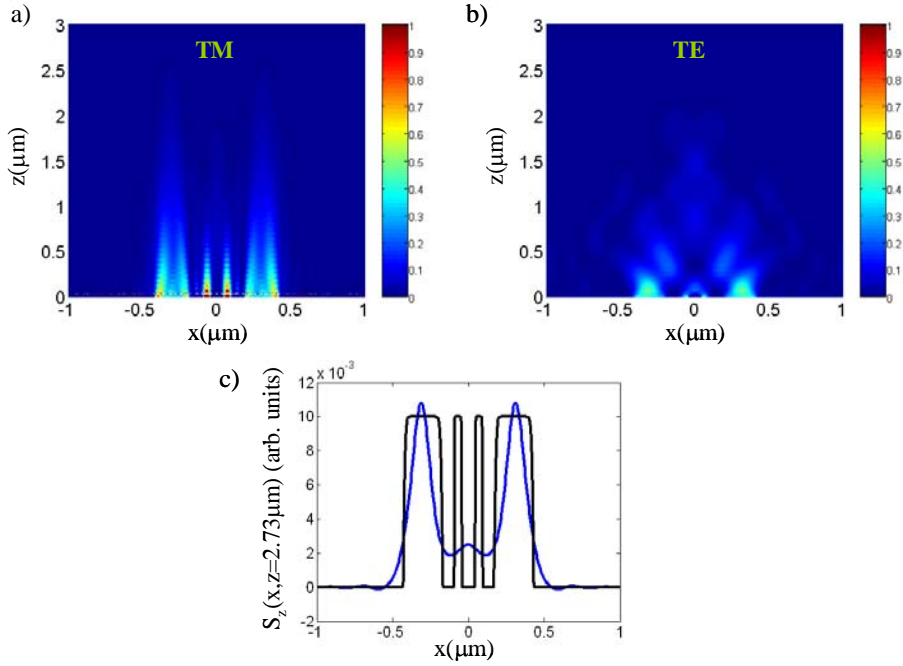

Fig. 12. Diffraction with a metallo-dielectric lens whose number of periods is eight times greater than the lens described in Fig.5. a) TM polarization. b) TE polarization. c) Section of $S_z$ at z=2.73μm, i.e. few nanometers beyond the end of the lens. Superimposed the position of the slits.

Of course, increasing the number of periods of the lens will affect the overall transmittance of the propagating modes and therefore the lens transparency, nevertheless the canalization mechanism endures such challenging trial. The consequences of these findings can be far reaching, because they might open the door to the examination of new optical phenomena: for example, it would be very intriguing to study how the metals and/or semiconductors cubic nonlinearities could affect the canalization mechanism in the framework of the formation of spatial, 1-D optical solitons. Although the sturdiness of the canalization mechanism has been studied for a particular metallo-dielectric lens (Ag/GaP), the results reported have a general validity and apply as well to different combination of metals and semiconductors.

## 4. Conclusions

In conclusion, we have discussed some important properties of the transmission function in the k-space for multilayered structures. In particular we have shown that the amplification of the evanescent modes can be directly related to the guided modes supported by the structure. The amplification of the evanescent mode is possible in generic structure both for TE and TM polarization. We have also shown that the canalization mechanism is independent of the excitation of plasmonic modes, but it takes place in an effective way only for TM polarization in metallo-dielectric structure thanks to the balance between the negative refraction of the Poynting vector in the metal layers and the positive refraction in the dielectric layers. We have also shown that the effectiveness of the canalization mechanism can be directly estimated by calculating the concavity of the phase of the transmission function at $k_x=0$. We hope that our results may be of some help for further development of more effective metallo-dielectric flat lens.

**Acknowledgments**
N. M. and G. D. thank the National Research Council for financial support.